\documentclass[12pt, a4paper]{article}

\usepackage[dvips]{graphicx}
\usepackage{wrapfig}
\usepackage{amssymb}
\usepackage{amsmath}
\usepackage[a4paper,hmargin={3cm,.9in},height=10in]{geometry}

\begin{document}

\title{Application of the EWL protocol to decision problems with imperfect
recall}

\author{\textsc{Piotr Fr\c{a}ckiewicz} \\Institute of Mathematics of the Polish Academy of Sciences\\
00-956, Warsaw, Poland}
\newtheorem{lemma}{Lemma}
\newtheorem{definition}{Definition}[section]
\newtheorem{theorem}{Theorem}
\newtheorem{proposition}{Proposition}[section]
\newtheorem{example}[proposition]{Example}
\newenvironment{proof}{\noindent\textit{Proof.}}
{\nolinebreak[4]\hfill$\blacksquare$\\\par}

\maketitle

\begin{abstract}
We investigate implementations of the Eisert-Wilkens-Lewenstein
scheme of playing quantum games beyond strategic games. The scope
of our research are decision problems, i.e., one-player extensive
games. The research is based on the examination of their features
when the decision problems are carried out via the EWL protocol.
We prove that unitary operators can be adapted to play the role of
strategies in decision problems with imperfect recall.
Furthermore, we prove that unitary operators provide the decision
maker possibilities that are inaccessible for classical
strategies.
\end{abstract}

\maketitle
\section{Introduction}
We present two new applications of the Eisert-Wilkens-Lewenstein
(EWL) protocol \cite{eisert}. The subject of applications are
decision problems with imperfect recall. Two studied applications
correspond to two main issues concerning such problems. The former
deals with the problem of no outcome-equivalence between mixed and
behavioral strategies that arises in games with imperfect recall.
We prove that extending the set of actions to unitary operators
may remove this non-equivalence. The latter part of our paper
concerns the problem of payoff maximization in the well-known
decision problem called the paradox of absentminded driver. We
reexamine the unitary operators treated as actions in the EWL
scheme applied to the paradox. We show that application of general
unitary operators yields benefit to the decision maker. The
strategy space $\mathsf{SU}(2)$ allows the decision maker to get
payoffs that are inaccessible by any classical strategy. We also
study the generalized EWL protocol extended to more than two
qubits and we demonstrate that some decision problems can be
treated by this generalized scheme.

\section{Preliminaries to game theory}
Definitions in this section are derived from \cite{osborne} and
\cite{rubinstein}. Readers who are not familiar with game theory
are encouraged to get acquainted with these books. The main object
that we are interested in is a decision problem. It is based on
the formal definition of the extensive game \cite{osborne} where
only one player acts. We restrict this term as much as it is
sufficient to be within the scope of our study.
\begin{definition}
A decision problem is a triple $\Gamma = \langle H, u, I\rangle$
where:
\begin{enumerate}
\item $H$ is a finite set of sequences called histories that
satisfies the following two properties:
\begin{enumerate}
\item The empty sequence $\emptyset$ is a member of $H$. \item If
$(a_k)_{k = 1,2,\dots, K} \in H$ and $K>1$ then $(a_k)_{k =
1,2,\dots, K-1} \in H$.
\end{enumerate}
A history $(a_k)_{k = 1,2,\dots, K} \in H$ is interpreted as a
feasible sequence of actions taken by the decision maker. The
history $(a_{1},a_{2},\dots,a_{K}) \in H$ is terminal if there is
no $(a_{1},a_{2},\dots,a_{K},a) \in H$. The sets of nonterminal
and terminal histories are denoted by $D$ and $Z$ respectively.
The set of actions available to the decision maker after
a~nonterminal history $h$ is defined by $A(h) = \{a\colon (h,a)
\in H\}.$ \item $u\colon Z \to \mathbb{R}$ is a utility function
which assigns a number (payoff) to each of the terminal histories.
\item The set of information sets, which is denoted by $I$, is a
partition of $D$ with the property that for all $h$, $h'$ in the
same cell of the partition $A(h) = A(h')$. Every information set
$I_{i}$ of the partition corresponds to the state of decision
maker's knowledge. When the decision maker when makes move after
certain history $h$ belonging to $I_{i}$, she knows that the
course of events of the decision problem takes the form of one of
histories being part of this information set. She does not know,
however, if it is the history $h$ or the other history from
$I_{i}$.
\end{enumerate}
\label{def1}
\end{definition}
The main method for describing decisions taken by a decision maker
is based on planning actions before she starts with her first
move. Every such plan is called a pure strategy:
\begin{definition}
A pure strategy $s$ is a function which assigns to every history
$h \in D$ an element of $A(h)$ with the restriction that if $h$
and $h'$ are in the same information set, then $s(h) = s(h')$.
\label{purestrategy}
\end{definition}
Let us denote by $e(h)$ {\it experience} of the decision maker. It
is the sequence of information sets and actions of the decision
maker along the history $h$. According to \cite{rubinstein}, a
decision problem has {\it imperfect recall} if there exists an
information set that contains histories $h$ and $h'$ for which
$e(h) \ne e(h')$ i.e., a decision maker forgets some information
about the succession of the information sets and (or) some of her
own past moves that she knew earlier.\\ The strategy set of a
decision maker can be extended to random strategies. There are two
ways of randomizing. One of them, known from strategic games,
specifies probability distribution over the set of pure strategies
and is called {\it mixed strategy}. The other specifies
probability distribution over the actions available to decision
maker at each information set:
\begin{definition}
A behavioral strategy $b$ is a function which assigns to every
history $h \in D$ a probability distribution $b(h)$ over $A(h)$
such that $b(h) = b(h')$ for any two histories $h$ and $h'$ which
belong to the same information set.
\end{definition}
Since different randomization of strategies may imply the same
utility payoff, a~possibility to measure what result particular
strategy produces is required:
\begin{definition}
Let mixed or behavioral strategy $\sigma$ in a decision problem be
given. The outcome $O(\sigma)$ of $\sigma$ is the probability
distribution over the terminal histories induced by $\sigma$. If
two different strategies $\sigma$ and $\sigma'$ induce the same
outcome then they are outcome-equivalent.
\end{definition}
The behavioral and mixed strategy ways of randomization are
outcome-equivalent in decision problems (more generally in
extensive games) with perfect recall. In problems with imperfect
recall some outcomes may be obtained only through a mixed strategy
or only through a behavioral strategy (see \cite{osborne} and
\cite{myerson}). This issue will be studied in Section~4.
\section{EWL scheme for quantum $2 \times 2$ strategic game}
The generalized Eisert-Wilkens-Lewenstein scheme \cite{eisert} is
defined by the following components:
\begin{enumerate}
\item an entangling operator $J$ composed of the identity operator
$I$ and Pauli operator~$\sigma_{x}$:
\begin{eqnarray}
J = \frac{1}{\sqrt{2}}(I \otimes I + i\sigma_{x} \otimes
\sigma_{x}),
\end{eqnarray}
\item unitary operators $U_{j}$, $j=1,2$, from the space
$\mathsf{SU}(2)$ of the form \cite{nawaz}:
\begin{eqnarray}
U_{j}(\theta, \alpha, \beta) = \cos\frac{\theta_j}{2}A_{j} +
\sin\frac{\theta_j}{2}B_{j} \quad  \mbox{for} \quad \theta_{j} \in
[0, \pi], \label{operator}
\end{eqnarray}
where $A_{j}$ and $B_{j}$ are defined as follows:
\begin{eqnarray}
A_{j}|0\rangle &= e^{i\alpha_{j}}|0\rangle, ~~~~~~~
A_{j}|1\rangle& = e^{-i\alpha_j}|1\rangle; \nonumber\\
B_j|0\rangle &= e^{i(\frac{\pi}{2} -\beta_{j})}|1\rangle, ~
~B_j|1\rangle& = e^{i(\frac{\pi}{2} +\beta_{j})}|0\rangle, ~~
\mbox{for} \quad \alpha, \beta \in [0,2\pi). \nonumber
\end{eqnarray}
\item a payoff function $\mathrm{E}(u)$ defined as expected value
of a discrete random variable $u$ with the values $\{u_{kl} \in
\mathbb{R}^2 \colon k,l = 0,1\}$ being payoffs associated with the
outcomes of a classical bimatrix $2 \times 2$ game, and the
probability distribution $p_{kl}$ defined by
$|\langle\psi_{f}|kl\rangle|^2$ where $|\psi_{f}\rangle =
J^{\dag}(U_{1} \otimes U_{2})J|00\rangle$ and $\{|kl\rangle\}_{k,l
\in \{0,1\}}$ is the computational base of $\mathbb{C}^2 \otimes
\mathbb{C}^2$:
\begin{eqnarray} \mathrm{E}(u)(U_{1} \otimes U_{2}) = \displaystyle \sum_{k,l \in \{0,1\}}
u_{kl}|\langle\psi_{f}|kl\rangle|^2. \label{dupa} \end{eqnarray}
As operators $U_{j}$ depend on parameters $\theta_{j}, \alpha_{j},
\beta_{j}$ we will sometimes denote payoff function
$\mathrm{E}(u)(U_{1} \otimes U_{2})$ as $\mathrm{E}(u)(\theta_{1},
\alpha_{1}, \beta_{1},\theta_{2}, \alpha_{2}, \beta_{2})$. If each
terminal history $h$ of a decision problem will be associated with
some outcome $o_{h}$ instead of some value of $u(h)$ we will write
$\mathrm{E}(O)$.
\end{enumerate}
In the EWL protocol two players select local operators from
(\ref{operator}) and each of them act on their own qubit initially
prepared in the $|0\rangle$ state. For a more detailed description
we encourage the reader to get acquainted with the prototype of
the EWL scheme in \cite{eisert} and other papers, for example
\cite{flitney0}, \cite{flitney} where authors have investigated
properties of the EWL protocol and compared them with the
classical $2\times2$ game.
\section{Decision problems with imperfect recall via EWL scheme}
The EWL scheme devised initially for a symmetric game Prisoner's
Dilemma has already been used for $2 \times 2$ games with
different properties in \cite{flitney0} and \cite{eisert2}, and
games with a~bigger number of strategies available to players
\cite{du}. Let us consider player's knowledge in a~strategic game
at the moment of taking an action. It is the same as in the case
of extensive games the property of which is that, players do not
have a possibility to watch the opponents' move. Another
similarity manifests itself, for instance, in a decision problem
when a decision maker has forgotten the actions chosen in some of
previous stages. Thus, such examples indicate a possibility of
applying the EWL protocol to this type of games as well. Our aim
is to adjust the EWL scheme to quantize two well-known decision
problems with imperfect recall.
\subsection{Application 1}
\begin{figure}[t]
\centering
\includegraphics[angle=0, scale=0.9]{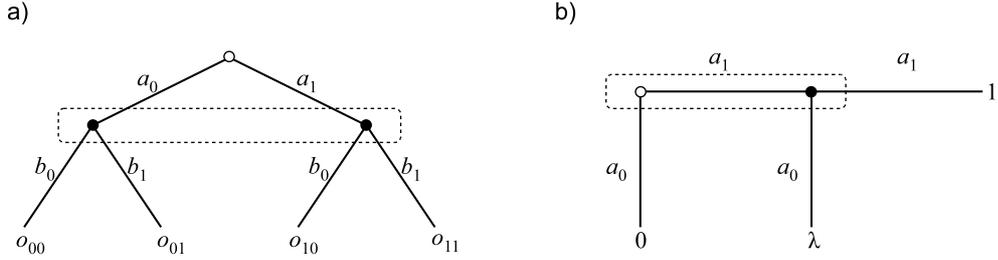}
\caption{Examples of decision problems with imperfect recall.}
\label{figure1}
\end{figure}

The first example is taken from \cite{osborne}. A decision maker
is faced with a choice between two possibilities. When she makes a
move, she has a choice of two actions once more. A significant
feature of this problem is that before taking another action the
decision maker forgets what action she has chosen previously.
Therefore, this problem exhibits imperfect recall. The formal
description $\langle H, O, I \rangle$ of this example according to
Definition \ref{def1} (with a small substitution of the payoff
function $u$ by an outcome function $O$) is as follows:
\begin{eqnarray}
&&H = \{\emptyset, a_{0}, a_{1}, (a_{0},b_{0}), (a_{0},b_{1}),
(a_{1},b_{0}), (a_{1},b_{1})\}; \nonumber\\ &&O(a_{k}, b_{l}) =
o_{kl},~~ \mbox{where} ~~ k,l \in \{0,1\}, ~~ I = \{\emptyset,
\{a_{0},a_{1}\}\}. \label{example1}
\end{eqnarray}
The decision problem in a `tree' language is shown in Figure
\ref{figure1}a.
As we have mentioned in preliminaries, the decision maker has two
different ways to precise her decision - expressed as mixed
strategies or as behavioral strategies. Her set of pure strategies
is $\{a_{0}b_{0}, a_{0}b_{1}, a_{1}b_{0}, a_{1}b_{1}\}$ where the
first (second) entry of $a_{k}b_{l}$ means an action taken by the
decision maker when she is in the first (second) information set.
Thus, she can choose mixed strategy as a probability distribution
over $a_{k}b_{l}$. On the other hand the decision maker can
specify an independent probability measure over actions available
at each information set i.e., her behavioral strategy is of the
form $((p,1-p),(q,1-q))$. This example shows no
outcome-equivalence between mixed and behavioral strategies. To be
precise, there are outcomes induced by some mixed strategies that
are not achievable by any behavioral strategy. To see this, let us
consider an outcome of the form $p_{1}o_{00} + (1-p_{1})o_{11}$,
$p \in (0,1)$. This outcome is obtained from a mixed strategy
$(p_{1}, 0, 0,1-p_{1})$. However, no behavioral strategy can yield
this outcome. To see this, notice that any behavioral strategy
$((p,1-p),(q,1-q))$ must assign probability $p(1-q)$ equal 0 to
yield the outcome $p_{1}o_{00} + (1-p_{1})o_{11}$. It implies that
the
probability of obtaining either $o_{00}$ or $o_{11}$ is equal 0.\\
Let us take a look how the EWL scheme can be applied to the
problem described above. Looking at the game tree we can see that
this problem has the same structure as an extensive form of a $2
\times 2$ strategic game. The decision maker before making her
first move takes on the role of the player 1 and afterwards takes
an action available for the player 2. As she has forgotten the
action taken previously, she has the same knowledge of the game as
players in $2 \times 2$ game. It is therefore natural to adapt to
this problem the EWL scheme when the decision maker chooses some
unitary operator $U_{1}$ with which she acts on the first qubit
$|\varphi\rangle_{1}$ and subsequently applies a unitary operator
$U_{2}$ on the second qubit $|\varphi\rangle_{2}$. Then a payoff
is calculated through the formula (\ref{dupa}). The formal
description of this problem in a manner comparable
 to (\ref{example1}) is as follows:
\begin{eqnarray}
&&H' = \{\emptyset, U_{1}|\varphi\rangle_{1},
(U_{1}|\varphi\rangle_{1},U_{2}|\varphi\rangle_{2})\}; \nonumber\\
&&u' = \mathrm{E}(O)(U_{1} \otimes U_{2}), ~~ I' =
\{\{|\varphi\rangle_{1}\}, \{|\varphi\rangle_{2}\}\}.
\label{quantumex}\end{eqnarray} It should be emphasized that we do
not try to identify each component from Definition \ref{def1} with
components of (\ref{quantumex}). Specification (\ref{quantumex})
takes on the informal character. It is aimed at organizing our
deliberations. The main feature of the EWL scheme is that it
comprises corresponding classical game i.e., there exists a set of
unitary operators that yield the same outcomes as classical
strategies. That is, classical actions can be realized via $I$ and
$i\sigma_{x}$ as it has been shown, for example, in
\cite{eisert2}. The decision maker's strategies are not single
actions in the problem (\ref{quantumex}), however. They are plans
that describe what does the decision maker do in each of her two
information sets (see Definition \ref{purestrategy}), i.e., what
unitary action she performs on each of two qubits individually.
Therefore, her set of pure classical strategies can be described
as $\{I \otimes I, I \otimes i\sigma_{x}, i\sigma_{x} \otimes I,
i\sigma_{x} \otimes i\sigma_{x}\}$. Then a classical mixed
strategy can be obtained by
\begin{eqnarray}
\sqrt{p_{1}}I \otimes I + \sqrt{p_{2}}I \otimes i\sigma_{x} +
\sqrt{p_{3}}i\sigma_{x} \otimes I + \sqrt{p_{4}}i\sigma_{x}
\otimes i\sigma_{x} \label{mixedstrategy} \end{eqnarray} where
$|\sqrt{p_{i}}|^2$ are the values of some probability mass
function (notice that $\sqrt{p_{i}}$ can take complex values). In
general case a (pure) unitary strategy takes the form
$U_{1}(\theta_{1}, \alpha_{1}, \beta_{1}) \otimes
U_{2}(\theta_{2}, \alpha_{2}, \beta_{2})$. In games represented by
bimatrices the equivalence (with respect to outcomes that can be
achieved) between classical actions and some fixed unitary
operators  is sufficient to claim that a quantum realization
generalizes the classical game. In extensive games, particularly
in the decision problem (\ref{quantumex}), it seems natural to
find unitary strategies that realize classical behavioral
strategies, not only mixed strategies. Following \cite{eisert2},
we know that a unitary strategy $U_{1}(\theta_{1}, 0, 0) \otimes
U_{2}(\theta_{2}, 0, 0)$ must imitate some classical move of the
decision maker. This strategy corresponds  exactly to classical
behavioral strategy $((p,1-p),(q,1-q))$. If we assume $p \equiv
\cos^2(\theta_{1}/2)$ and $q \equiv \cos^2(\theta_{1}/2)$, we
obtain from (\ref{dupa}):
\begin{eqnarray}
\mathrm{E}(O)(\theta_{1}, \theta_{2}) = \sum_{k,l \in
\{0,1\}}o_{kl}\cos^2\left(\frac{\theta_{1} -
k\pi}{2}\right)\cos^2\left(\frac{\theta_{2} - l\pi}{2}\right).
\end{eqnarray}
Since one-parameter operators $U_{j}(\theta_{j},0,0)$ implement
classical moves, a natural question arises: what is the role of
wider range of unitary strategies in the decision
problem~(\ref{quantumex})? The answer to this question is
surprising: Extension of the set of behavioral strategies to the
set $\mathsf{SU}(2) \otimes \mathsf{SU}(2)$ causes
outcome-equivalence of behavioral strategies with mixed
strategies. Notice that this problem is not trivial because there
is no identity of the expression (\ref{mixedstrategy}) and
$U_{1}(\theta_{1},\alpha_{1},\beta_{1}) \otimes
U_{2}(\theta_{2},\alpha_{2},\beta_{2})$. For example, if one puts
$\sqrt{p_{1}} = \sqrt{p_{4}} = \frac{1}{\sqrt{2}}$ and
$\sqrt{p_{2}} = \sqrt{p_{3}} = 0$, there is no representation of
(\ref{mixedstrategy}) in the form of the tensor product of
(\ref{operator}). On the other hand when we take, for example,
$\alpha_{1} = \alpha_{2} \ne 0$ then the tensor product
$U_{1}(\theta_{1}, \alpha_{1}, \beta_{1}) \otimes
U_{2}(\theta_{2}, \alpha_{2}, \beta_{2})$ has not the form of a
mixed strategy for any angles $\theta_{1}, \theta_{2}, \beta_{1},
\beta_{2}$. However, the following statement is true:
\begin{proposition} For any mixed strategy of a decision maker in
the decision problem (\ref{quantumex}) there is an
outcome-equivalent pure unitary strategy. \label{proposition0}
\end{proposition}
\begin{proof} The set of outcomes yielded by all mixed strategies is a convex hull
of elements $\{o_{00}, o_{01}, o_{10}, o_{11}\}$ due to the
expression for a mixed strategy (\ref{mixedstrategy}) or,
equivalently, a mixed strategy of the decision problem
(\ref{example1}). We will prove that any convex combination
$\sum_{k,l \in \{0,1\}}p_{kl}o_{kl}$ can be written as an expected
outcome $\mathrm{E}(O)(U_{1} \otimes U_{2}) = \sum_{k,l \in
\{0,1\}} o_{kl}|\langle\psi_{f}|kl\rangle|^2$ for some unitary
operations $U_{1}$ and $U_{2}$ from $\mathsf{SU}(2)$. At first let
us consider the case $p_{00} = p_{11} = 0$ or $p_{01} = p_{10} =
0$. Then the convex combination $\sum_{k,l \in
\{0,1\}}p_{kl}o_{kl}$ is a segment $p_{01}o_{01} + p_{10}o_{10}$
or $p_{00}o_{00} + p_{11}o_{11}$, respectively. Putting
$U_{1}(0,\alpha_{1},0) \otimes U_{2}(0,0,0)$ we get $\mathrm{E}(O)
= o_{00}\cos^2\alpha_{1} + o_{11}\sin^2\alpha_{1}$ that is a
segment linking points $o_{00}$ and $o_{11}$. Similarly, if we
take $U_{1}(\pi,0,\beta_{1}) \otimes U_{2}(0,0,0)$ we obtain
$o_{01}\sin^2\beta_{1} + o_{10}\cos^2\beta_{1}$. Now, let us
examine general convex combination of points $o_{kl}$ such that
$p_{00} + p_{11} \ne 0$ and $p_{01} + p_{10} \ne 0$. The
combination $\mathrm{E}(O)$ associated with
$U_{1}(\theta_{1},\alpha_{1},\beta_{1}) \otimes U_{2}(0,0,0)$ is
of the form:
\begin{eqnarray}
\left(o_{00}\cos^2\alpha_{1} +
o_{11}\sin^2\alpha_{1}\right)\cos^2\frac{\theta_{1}}{2} +
\left(o_{01}\sin^2\beta_{1} +
o_{10}\cos^2\beta_{1}\right)\sin^2\frac{\theta_{1}}{2}.
\label{combination}
\end{eqnarray}
Comparing the coefficients of the combination $\sum_{k,l \in
\{0,1\}}p_{kl}o_{kl}$ and (\ref{combination}) we obtain the system
of equations that has a unique solution:
\begin{eqnarray}
\cos^2\frac{\theta_{1}}{2} = p_{00} + p_{11}, ~~\cos^2\alpha_{1} =
\frac{p_{00}}{p_{00} + p_{11}}, ~~\cos^2\beta_{1} =
\frac{p_{10}}{p_{01} + p_{10}}. \label{rozwiazanie}
\end{eqnarray}
The result (\ref{rozwiazanie}), together with the first case,
finishes the proof.
\end{proof}
Notice that the unitary strategies used in the proof depend only
on an operation on the first qubit.  Due to the fact that the
qubits are maximally entangled every outcome can be obtained by
performing an operation only on the first or only on the second
qubit.
\subsection{Application 2} The next example in which we are going to use
the EWL scheme is based on \cite{rubinstein}. As the previous
example, this one also shows difference between reasoning based on
mixed and behavioral strategies. The application is dealing with
the well-known imperfect recall problem called the paradox of
absentminded driver. Our research is not the first attempt to put
this problem into quantum domain. The first one appeared in
\cite{cabello}. The authors of this paper presented the way of
quantization with the use of the Marinatto and Weber scheme of
playing quantum $2 \times 2$ games \cite{marinatto} - the initial
state plays the main role. In outline, for many kinds of the
absentminded driver problems various initial states are chosen to
maximize the driver's payoff. Therefore, we expect that no other
protocol could be ahead of \cite{cabello} in terms of maximization
of the driver's payoff. However, the quantum version based on the
EWL protocol turns out to be a convenient way to make an analysis
of some complicated cases of the problem of absentminded driver.
\subsubsection{The paradox of absentminded driver.} The name of this decision
problem is derived from a certain story describing this issue. An
individual sitting for some time in a pub eventually decides to go
back home. The way is leading through the motorway with two
subsequent exits. The first exit leads to a catastrophic area
(payoff 0). The choice of the other one will lead the decision
maker home (payoff $\lambda > 2$). If he continues his journey
along the motorway not choosing any of the exits, he will not be
able to go back home but he has a possibility to stay for the
night at a motor lodge (payoff 1). The key determinant is the
driver's absent-mindedness. This means that when he arrives at the
exit he is not able to tell if it is the first or the second exit
due to his absent-mindedness. This situation is described on
Figure \ref{figure1}b. The formal description is as follows:
\begin{eqnarray}
&&H = \{\emptyset, a_{0}, a_{1}, (a_{1},a_{0}), (a_{1},a_{1}),\},
~~ I = \{\emptyset, a_{1}\}; \nonumber\\ &&u(a_{0}) = 0,~~
u(a_{1}, a_{0}) = \lambda, ~~ u(a_{1}, a_{1}) = 1.
\label{example2}
\end{eqnarray}
Let us determine decisions that the driver can make. Since the
decision maker has just one information set, according to
Definition \ref{purestrategy}, only two pure strategies are
available to him: `exit' or `motorway' with respective terminal
histories $a_{0}$ and $(a_{1},a_{1})$. Similarly, behavioral
strategy of the driver will be represented by the same random
device in each of the two nodes of the information set i.e., it is
on the form $(p,1-p)$ where $p$ is the probability of choosing
`exit'. Notice first that the driver plans his journey still
sitting in the bar which is equivalent to choosing some pure
strategy. The optimal strategy is `motorway' (with corresponding
payoff 1) which becomes paradoxical when the decision maker begins
carrying out this pure strategy. It is better for him, when he
approaches an exit, to go away from the motorway because he comes
to a conclusion that with equal probability he is at the first or
the second exit. Consequently, his optimal choice will be a
certain behavioral strategy. For example if $\lambda = 4$, the
expected payoff corresponding to the strategy $(p,1-p)$ is
expressed by $u(p,1-p) = (1-p)(1+3p)$. Maximizing $u(p,1-p)$ we
conclude that the optimal decision for the decision maker is to
choose `exit' with probability $1/3$ each time he encounters an
intersection, which corresponds to the expected payoff $4/3$. As
in the previous example, here as well we can notice lack of
equivalence between behavioral and mixed strategies. This time,
however, behavioral strategy is strictly better than mixed one as
it ensures strictly higher payoff for the driver. Observe,
however, that condition $\lambda > 2$ is essential for this case.
Otherwise, $p=0$ maximizes the expected payoff $u(p,1-p)$ which is
equal 1.

Now, we are going to implement the EWL protocol to this problem.
It is possible since in the classical example we have again a
decision problem with two stages. Moreover, actions are taken
independently at each of these stages as in the $2 \times 2$
bimatrix game. Further, each player in the $2 \times 2$ game has
not any knowledge of an action taken by his opponent. Therefore,
this is the same situation as if the decision maker was in the
role of the player 1 and then the player 2, and he forgot his
previous move. Let us assign the state after each action of the
classical decision problem with the computational base of
respective qubit. States induced by actions: `exit' and `motorway'
available after history $\emptyset$ correspond to $|0\rangle$ and
$|1\rangle$ states of the first qubit $|\varphi\rangle_{1}$.
Similarly, we assign states after actions from $A(a_{1})$ (see
Definition \ref{def1}) to base states of the second qubit
$|\varphi\rangle_{2}$. Notice that this is the obvious procedure
applied in quantum $2 \times 2$ bimatrix games where outcomes are
assigned to base states $|kl\rangle$ where $k,l \in \{0,1\}$. We
assume, as in the classical case, that in the quantum realization
the driver is unable to distinguish to which qubit he applies a
unitary action. Therefore, the two qubits are in the information
set. It implies that the same unitary operation $U$ is applied to
both qubits. More formally:
\begin{eqnarray}
&&H' = \{\emptyset, U|\varphi\rangle_{1},
(U|\varphi\rangle_{1},U|\varphi\rangle_{2})\}, ~~ I' =
\{|\varphi\rangle_{1}, |\varphi\rangle_{2}\}; \nonumber\\
&&\mathrm{E}(u)\left(U^{\otimes 2}\right) =
\lambda|\langle\psi_{f}|10\rangle|^2 +
1|\langle\psi_{f}|11\rangle|^2 ~~\mbox{and}~~ |\psi_{f}\rangle =
J^{\dag}U^{\otimes2}J|00\rangle.\label{quantumdp1}\end{eqnarray}
The core of the issue lies in the payoff function $\mathrm{E}(u)$.
If state $|0\rangle$ on the first qubit is measured (which
corresponds to choosing `exit' at the first intersection), the
payoff assigned to this state equals 0 regardless of the state
measured on the second qubit. Therefore, in (\ref{quantumdp1}) the
expected payoff $\mathrm{E}(u)$ includes
$0\left(|\langle\psi_{f}|00\rangle|^2 +
|\langle\psi_{f}|01\rangle|^2\right)$. Thus, so defined quantum
realization generalizes the classical case. The classical pure
strategies can be again implemented by $I$ and $i\sigma_{x}$ which
correspond to `exit' and `motorway', respectively.  These
strategies imply operations $I^{\otimes 2}$ and
$(i\sigma_{x})^{\otimes 2}$ on both qubits that are
indistinguishable by the decision maker, and produce outcomes
equal to classical ones. Although we have assumed that the result
0 on the first qubit determine payoff 0, we always have to specify
operations on both qubits. Like in the classical case, also here
decision maker's strategy have to precise an action in every
possible state of a decision problem. As in the previous example
one-parameter operation $U(\theta, 0, 0)$ matches classical
behavioral strategy and we have:
\begin{eqnarray}
\mathrm{E}(u)(\theta) =
\lambda\sin^2\frac{\theta}{2}\cos^2\frac{\theta}{2} +
1\sin^4\frac{\theta}{2}. \label{oneparameterpayoff}
\end{eqnarray}
If we replace $\cos^2(\theta/2)$ by $p$ in the formula
(\ref{oneparameterpayoff}) we obtain expected payoff that
corresponds to the classical behavioral strategy $(p,1-p)$ in the
decision problem (\ref{example2}). From the classical case
(\ref{example2}) we already know that the driver can obtain the
maximal payoff which is $4/3$ when $\lambda = 4$, using operators
of the type $U(\theta,0,0)$. Let us investigate if the decision
maker can benefit when the range of his actions is extended to any
operator of the form (\ref{operator}). Assume that the driver has
two-parameter set of unitary operations $U(\theta, \alpha, 0)$ at
his disposal. Then the expected payoff is as follows:
\begin{eqnarray}
\mathrm{E}(u)(\theta, \alpha) = \frac{1}{4}\lambda(\sin2\alpha +
1)\sin\theta + \left(\sin(2\alpha)\cos^2\frac{\theta}{2} -
\sin^2\frac{\theta}{2}\right)^2
\end{eqnarray}
If the driver applies $U(\pi/2, \pi/4, 0)$ to his both qubits, the
expected payoff equals $\lambda/2$. Coming back to the case
$\lambda = 4$, he gets 2 utilities instead of $4/3$. Moreover,
this is the highest payoff which the driver can guarantee himself
by using any unitary operations (\ref{operator}). To prove this,
we determine the final state $|\psi_{f}\rangle$ when any unitary
operation $U(\theta, \alpha, \beta)^{\otimes2}$ is given. The
matrix representation of $|\psi_{f}\rangle$ takes the form:
\begin{eqnarray} |\psi_{f}\rangle = \left(\begin{array}{c}
\displaystyle\cos2\alpha\cos^2\frac{\theta}{2}
+ \sin2\beta\sin^2\frac{\theta}{2} \\
\displaystyle
i\left(\cos\frac{\theta}{2}\sin\frac{\theta}{2}(\cos(\alpha -
\beta) + \sin(\alpha - \beta)\right) \\
\displaystyle
i\left(\cos\frac{\theta}{2}\sin\frac{\theta}{2}(\cos(\alpha
- \beta) + \sin(\alpha - \beta)\right)\\
\displaystyle\sin2\alpha\cos^2\frac{\theta}{2} -
\cos2\beta\sin^2\frac{\theta}{2}
\end{array}\right). \end{eqnarray}
Therefore, the state $|\psi_{f}\rangle$ is a particular case of
the state $|\psi'_{f}\rangle$ of the form:
\begin{eqnarray}
|\psi'_{f}\rangle = \sum_{k,l \in \{0,1\}}\eta_{kl}|kl\rangle, ~
\mbox{where} ~ \eta_{kl} \in \mathbb{C}, \sum_{k,l \in
\{0,1\}}|\eta_{kl}|^2 = 1~ \mbox{and} ~ \eta_{01} = \eta_{10}.
\end{eqnarray}
Exchanging $|\psi_{f}\rangle$ by $|\psi'_{f}\rangle$ in
(\ref{quantumdp1}), we see that the expected payoff
$\mathrm{E}'(u)$ equals $\lambda|\eta_{10}|^2 + |\eta_{11}|^2$.
Furthermore, we have $\lambda
> 2$. It implies that the set
$\mathrm{arg}\max_{\eta_{kl}}(\mathrm{E}'(u))$ consists of all
points $(\eta_{10},\eta_{11})$ for which
$(|\eta_{10}|^2,|\eta_{11}|^2) = (1/2,0)$. It is obvious that in
the special case $|\psi_{f}\rangle$ the equality
$\mathrm{arg}\max_{\psi_{f}}(\mathrm{E}(u)(\theta,\alpha,\beta)) =
\mathrm{arg}\max_{\psi_{f}}(|\langle\psi_{f}|10\rangle|^2)$ is
fulfilled as well. As we obtain
$\max_{\psi_{f}}(|\langle\psi_{f}|10\rangle|^2) = 1/2$, the
decision maker can achieve maximal payoff equal to $\lambda/2$.
Observe that the maximal payoff that the decision maker can get in
the classical case is $\lambda^2/4(\lambda - 1)$. This is strictly
less than $\lambda/2$ if only $\lambda > 2$. This leads us to the
conclusion that in the decision problem (\ref{example2}) extended
to the quantum domain there exists the unitary strategy for any
$\lambda
> 2$ that is strictly better than any classical one.
\subsubsection{The n-tuple paradox of absentminded driver.} 
\begin{figure}[t]
\centering
\includegraphics[angle=0, scale=0.9]{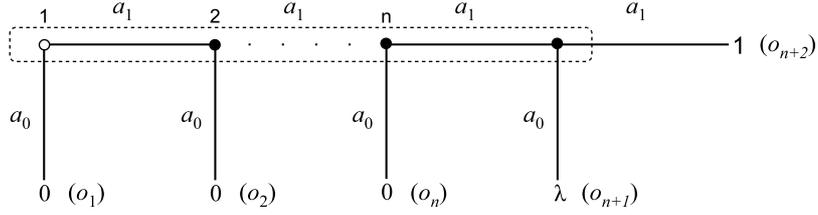}
\caption{The n-tuple decision problem of absentminded driver.}
\label{figure2}
\end{figure}

We have already showed the advantage of quantum strategies over classical ones in the
problem of absentminded driver. Now, we test unitary strategies in
the case where the driver comes across more than one treacherous
intersection. At this moment we make an assumption that the
absentminded driver problem is  characterized by $n+1$
intersections such that the first
 $n$ intersections are treacherous ones (payoff 0 when `exit' is chosen at each of these intersections),
and only one action `exit' taken at n+1 intersection leads the
driver home (payoff $\lambda$). Choosing the action `motorway' all
the time yields payoff 1, the same as in (\ref{example2}). This
problem is depicted on Figure \ref{figure2}.
A formal description of this classical case is similar to
(\ref{example2}). To find an optimal classical strategy we have to
maximize $(1-p)^n((\lambda-1)p+1)$ In quantum version we apply the
general EWL protocol for $(n+1) \times (n+1)$ bimatrix game where
entangling operator $J$ and final state $|\psi_{f}\rangle$ take
the form given in \cite{benjamin}:
\begin{eqnarray}
J = \frac{1}{\sqrt{2}}(I^{\otimes n+1} + i\sigma_{x}^{\otimes
n+1}), ~~ |\psi_{f}\rangle = J^{\dag}U^{\otimes
n+1}J|0\rangle^{\otimes n+1} \label{nEWL}
\end{eqnarray} The decision maker
carries out some fixed unitary operation $U \in \mathsf{SU}(2)$ on
each of $n + 1$ qubits. In addition, the qubits belong to the same
information set:
\begin{eqnarray}
&&H' = \{\emptyset, U|\varphi\rangle_{1},
(U|\varphi\rangle_{1},U|\varphi\rangle_{2}),\dots,(U|\varphi\rangle_{1},U|\varphi\rangle_{2},\dots,U|\varphi\rangle_{n+1})\}; \nonumber\\
&&I' =
\{|\varphi\rangle_{1}, |\varphi\rangle_{2},\dots,|\varphi\rangle_{n+1}\}; \nonumber\\
&&\mathrm{E}(u)\left(U^{\otimes n+1}\right) =
\lambda|\langle\psi_{f}|1\rangle^{\otimes n}|0\rangle|^2 +
|\langle\psi_{f}|1\rangle^{\otimes
n+1}|^2.\label{quantumdp2}\end{eqnarray} We identify the states
after actions `exit' and `motorway' of the decision problem from
Figure \ref{figure2} with states $|0\rangle$ and $|1\rangle$,
respectively, exactly like in (\ref{quantumdp1}). This causes the
equivalence between the decision problem (\ref{quantumdp2}) with
unitary operators reduced to one-parameter operators
$U(\theta,0,0)$ and the classical case, in the same way as in
(\ref{example2}) and (\ref{quantumdp1}):
\begin{eqnarray}
\mathrm{E}(\theta) =
\lambda\cos^2\frac{\theta}{2}\sin^{2n}\frac{\theta}{2} +
\sin^{2(n+1)}\frac{\theta}{2}. \label{classicalE}
\end{eqnarray}
The $n$ payoffs equal 0 could suggest these ones are essential so
that the EWL scheme can generalize the decision problem depicted
in Figure \ref{figure2}, but this is not true. In fact, any
decision problem given by the decision tree depicted in Figure
\ref{figure2} can be implemented by the EWL scheme. (Notice that
(\ref{example2}) represents any decision problem given by the
decision tree shown in Figure \ref{figure1}b as it comes down to a
problem with payoffs $0, \lambda, 1$ through adding a respective
constant to all the payoffs and (or) multiplying all the payoffs
by a respective constant). To demonstrate that the EWL quantum
representation defines correctly any kind of the n-tuple paradox,
we assign, instead of fixed payoffs, outcomes $o_{1},
o_{2},\dots,o_{n+2}$ to the terminal histories. Then the n-tuple
decision problem takes the form:
\begin{eqnarray}
&&H' = \{\emptyset, a_{0}, a_{1},\dots,(a^1_{1},\dots,a^{n}_{1},
a^{n+1}_{0}), (a^1_{1},\dots, a^n_{1}, a^{n+1}_{1})\}; \nonumber\\
&&I' = \{\emptyset, a_{1},\dots, (a^1_{1},\dots, a^{n}_{1},
a^{n+1}_{1})\}; \nonumber\\
&&O(a_{0}) = o_{1}, ~~ O(a^1_{1},\dots,a^{t}_{1},a^{t+1}_{0}) =
o_{t+1} ~~ \mbox{for} ~~ t=1,2,\dots,n. \label{npado}
\end{eqnarray}
Let us denote by $|j_{1}, j_{2},\dots,j_{n+1}\rangle \in
\bigotimes_{n+1}\mathbb{C}^2$ an element of the computational
base. For any $t=1,2,\dots,n$ a symbol
 $(j_{t},j_{t+1},\dots,j_{n+1})_{2}$
denotes the binary representation of a~(decimal) number .
\begin{proposition}
The decision problem (\ref{quantumdp2}) with unitary operators $U$
restricted to $U(\theta,0,0)$ and expected payoff function
$\mathrm{E}(O)$ extended to include:
\begin{eqnarray}
\sum_{t=1}^no_{t}\sum_{x=0}^{2^{n-t+1}-1}|\langle\psi_{f}|1\rangle^{\otimes
t-1}|0\rangle|x\rangle|^2 ~~ \mbox{where} ~~ x =
(j_{t+1},j_{t+2},\dots,j_{n+1})_{2} \label{wzorproposition}
\end{eqnarray}
implements the decision problem (\ref{npado}).
\label{proposition2}
\end{proposition}
\begin{proof}
First we calculate $U(\theta,0,0)^{\otimes n+1}J|0\rangle^{\otimes
n+1}$ where $J$ is defined by (\ref{nEWL}). Then the expression
$\sqrt{2}U^{\otimes n+1}J|0\rangle^{\otimes n+1}$ takes the form:
\begin{eqnarray}
\sum_{y=0}^{2^{n+1}-1}\left(i^{r(y)}\cos^{n-r(y)+1}\frac{\theta}{2}\sin^{r(y)}\frac{\theta}{2}
 -
 i^{n-r(y)}\cos^{r(y)}\frac{\theta}{2}\sin^{n-r(y)+1}\frac{\theta}{2}\right)|y\rangle
 \label{Urazy0}
\end{eqnarray}
where the element $r(y)$ depends on $y =
(j_{1},j_{2},\dots,j_{n+1})_{2}$ and is given be the formula $r(y)
= j_{1} + j_{2} + \dots +j_{n+1}$. Let us fix any element
$|y\rangle$ from the computational base and determine the inner
product $\langle y|\psi_{f}\rangle$. To avoid laborious
computation that are necessary to obtain complete form of the
final state $|\psi_{f}\rangle$ we can choose the following simpler
way to calculate the inner product. We take the bra vector
 $J^{\dag}_{y,\cdot} = (\langle
y| - i\langle \overline{y}|)\sqrt{2}$ where $|\overline{y}\rangle
= \sigma_{x}^{\otimes n+1}|y\rangle$. In a language of matrices
the element $J^{\dag}_{y,\cdot}$ is the $y$-th row of a matrix
representation of $J^{\dag}$. Next, let us put $P$ as a label of a
projector $|y\rangle \langle y| + |\overline{y}\rangle \langle
\overline{y}|$. Then $\langle y|\psi_{f}\rangle$ can be expressed
as follows:
\begin{eqnarray}
\langle y|\psi_{f}\rangle = J^{\dag}_{y,\cdot}PU^{\otimes
n+1}J|0\rangle^{\otimes n+1}. \label{puj}
\end{eqnarray}
Notice that $r(y)$ and $r(\overline{y})$ are connected through
equation $r(y) +r(\overline{y}) = n+1$. Using this fact and a
result from (\ref{Urazy0}) the amplitude associated with
 $|\overline{y}\rangle$ of the state $U^{\otimes
n+1}J|0\rangle^{\otimes n+1}$ is given by
\begin{eqnarray}
\frac{1}{\sqrt{2}}\left(i^{r(\overline{y})}\cos^{r(y)}\frac{\theta}{2}\sin^{r(\overline{y})}\frac{\theta}{2}
-
i^{n-r(\overline{y})}\cos^{r(\overline{y})}\frac{\theta}{2}\sin^{r(y)}\frac{\theta}{2}\right).
\label{y_}
\end{eqnarray}
In order to complete the state $PU^{\otimes
n+1}J|0\rangle^{\otimes n+1}$ we just copy the amplitude of
$|y\rangle$ from (\ref{Urazy0}). If we use (\ref{puj}) and
(\ref{y_}) we will receive the final form of $\langle
y|\psi_{f}\rangle$:
\begin{eqnarray}
\langle y|\psi_{f}\rangle =
i^{r(y)}\cos^{r(\overline{y})}\frac{\theta}{2}\sin^{r(y)}\frac{\theta}{2}.
\label{iloczynskalarny}
\end{eqnarray}
The result (\ref{iloczynskalarny}) together with substitution $p =
\cos^2(\theta/2)$ immediately gives us $|\langle
\psi_{f}|y\rangle|^2 = p^{r(\overline{y})}(1-p)^{r(y)}$. After
some calculations we conclude from the last result that:
\begin{eqnarray}
\sum_{x}|\langle\psi_{f}|y\rangle|x\rangle|^2 =
p^{r(\overline{y})}(1-p)^{r(y)}\sum_{x} {r(x) + r(\overline{x})
\choose r(x)}p^{r(\overline{x})}(1-p)^{r(x)} \label{dwumian}
\end{eqnarray}
The sum on the right-hand side of the equation (\ref{dwumian}) is
the Newton's formula which is equal 1. Therefore, the formula
(\ref{dwumian}) leads us to a conclusion that for any
$t=1,2,\dots,n$ the component assigned to $o_{t}$ of the formula
(\ref{wzorproposition})
 can be expressed as:
\begin{eqnarray}
\sum_{x}|\langle\psi_{f}|1\rangle^{\otimes
t-1}|0\rangle|x\rangle|^2 = (1-p)^{t-1}p. \label{podstawienie}
\end{eqnarray}
Equation (\ref{podstawienie}) ends the proof as the right-hand
side of the equation is a probability of the outcome $o_{t}$ in
the decision problem (\ref{npado}) when a behavioral strategy
$(p,1-p)$ is taken.
\end{proof}
It follows from Proposition \ref{proposition2} that every time
when we concern the classical problem depicted on Figure
\ref{figure2} we can consider the problem (\ref{quantumdp2}) when
unitary operators (\ref{operator}) are restricted to one-parameter
operators $U(\theta,0,0)$ and the payoff function is given by
(\ref{classicalE}).

Let us return to example (\ref{quantumdp2}) where a payoff
function is fixed. Let us check if unitary operators
(\ref{operator}) can yield strictly better results than any
classical strategies applied to the problem. In order to do that,
we need to determine the expected payoff $\mathrm{E}(u)(U)$
defined in (\ref{quantumdp2}) for any $U(\theta,\alpha,\beta) \in
\mathsf{SU}(2)$ . We can find the components
$|\langle\psi_{f}|1\rangle^{\otimes n}|0\rangle|^2$ and
$|\langle\psi_{f}|1\rangle^{\otimes n+1}|^2$ of $\mathrm{E}(u)(U)$
with the use of equation (\ref{puj}). After simple calculations we
get:
\begin{eqnarray}
\mathrm{E}(u)(U)&=&
\lambda|i\cos^{n}\frac{\theta}{2}\sin\frac{\theta}{2}\sin(n\alpha
- \beta) +
i^n\cos\frac{\theta}{2}\sin^{n}\frac{\theta}{2}\cos(\alpha -
n\beta)|^2 \nonumber \\ &+&
1|\cos^{n+1}\frac{\theta}{2}\sin[(n+1)\alpha] +
i^{n+1}\sin^{n+1}\frac{\theta}{2}\cos[(n+1)\beta]|^2.
\label{3parameters}
\end{eqnarray}
Comparing the payoffs (\ref{classicalE}) and (\ref{3parameters})
we really ought to expect better results yielded by 3-parameter
operators.
\begin{example}
For $n=3$ and $\lambda = 20$ the maximization result of
(\ref{classicalE}) and (\ref{3parameters}) is as follows:
\begin{eqnarray}
&&\mbox{classical scenario:}~~\max_{\theta}{\mathrm{E}(u)(\theta)}
\approx
\mathrm{E}(u)\left(\frac{7\pi}{10}\right)  \approx 2,46;\nonumber\\
&&\mbox{EWL scenario:}~~\max_{\theta, \alpha,
\beta}{\mathrm{E}(u)(\theta, \alpha, \beta)} =
\mathrm{E}(u)\left(\frac{\pi}{2}, \frac{9\pi}{16},
\frac{3\pi}{16}\right) = 5. \nonumber
\end{eqnarray}
\end{example}
This example shows that an increased number of treacherous
intersections does not reduce the ability of the EWL scheme to
yield benefit to the decision maker. Moreover, the ratio
$\max_{\theta, \alpha, \beta}\mathrm{E}(u)(\theta, \alpha, \beta):
\max_{\theta}\mathrm{E}(u)(\theta)$ grows together with $\lambda$.
For large $n$ we had better use some mathematical software to
determine the precise result of optimization. Following
proposition assure us that the attempt of finding optimal solution
makes sense for any $n$:
\begin{proposition}
For any $n \geqslant 1$ in the n-tuple decision problem
(\ref{quantumdp2}) there exist a~number $\lambda_{0} >1$, angles:
$\theta' \in (0, \pi)$ and $\alpha', \beta' \in (0, 2\pi)$ such
that for any payoff $\lambda'
> \lambda_{0}$ the unitary strategy $U(\theta',\alpha',\beta')$ yields a payoff strictly
higher than a payoff achieved by any classical strategy.
\label{proposition3}
\end{proposition}
\begin{proof}
We showed in subsection 4.2.1 the case when $n=1$. So we now
assume $n \geqslant 2$. Let us take the decision problem
(\ref{quantumdp2}). For any $n \geqslant 2$ let us put
$\lambda_{0} =
(\cos^{2n}\frac{\theta'}{2}\sin^2\frac{\theta'}{2})^{-1}$ and
unitary operator $U^*(\theta', \alpha', \beta')$ defined by:
\begin{eqnarray}
\theta' = 2\arccos\frac{1}{\sqrt{n+1}},~~ \alpha' = \frac{(\pi +
2\pi\chi_{A}(n))n}{2(n^2 - 1)},~~ \beta' = \frac{\pi +
2\pi\chi_{A}(n)}{2(n^2 - 1)}. \label{parametersunitary}
\end{eqnarray}
where $\chi_{A}(n)$ is an indicator function of a set $A =
\{n\colon i^{n-1} = -1\}$. Notice that $\lambda_{0}, \theta',
\alpha'$ and $\beta'$ all meet the requirements of the
proposition. They depend only on $n$. Further, $\lambda_{0}$ is
well defined as $\theta' \notin \{0,\pi\}$. For any $\delta
> 1$ let us denote $\lambda' = \delta \lambda_{0}$ the payoff
associated with the problem (\ref{quantumdp2}).  By putting
parameters (\ref{parametersunitary}) into formula
(\ref{3parameters}) and comparing it to (\ref{classicalE}), and by
using the fact that $\theta' \in
\mathrm{arg\,max}_{\theta}\left(\cos^2\frac{\theta}{2}\sin^{2n}\frac{\theta}{2}\right)$,
we obtain the following sequence of inequalities:
\begin{eqnarray}
\mathrm{E}(u)(U^*) &\geqslant&
\lambda'\left(\cos^{2n}\frac{\theta'}{2}\sin^2\frac{\theta'}{2} +
\cos^2\frac{\theta'}{2}\sin^{2n}\frac{\theta'}{2} \right) = \delta
+
\lambda'\cos^2\frac{\theta'}{2}\sin^{2n}\frac{\theta'}{2} \nonumber\\
&>& 1 + \lambda'\cos^2\frac{\theta'}{2}\sin^{2n}\frac{\theta'}{2}
\geqslant \max_{\theta \in [0, \pi]}\mathrm{E}(u)(\theta),
\end{eqnarray}
which completes the proof.
\end{proof}
Let us observe now, how Proposition \ref{proposition3} concerns
the result of quantization the absentminded driver in subsection
4.2.1. A segment of numbers $\lambda$ in which there exists
unitary strategy strictly better than classical one, is the
segment $(2, \infty)$. It can be easily proved that for any
$\lambda \in \mathbb{R} \setminus (2, \infty)$ the maximal payoff
of the decision problem (\ref{example2}) is equal 1 regardless of
used unitary strategies. Proposition \ref{proposition3} shows that
the problem of finding optimal unitary strategy in decision
problems with various numbers of the treacherous intersections is
very much alike.
\section{Conclusion} We have found the new use of the EWL protocol
beyond strategic $2 \times 2$ games. It turns out once again that
game theory defined on quantum domain provides results that are
inaccessible in classical game theory. We have confirmed through
Proposition \ref{proposition3} that we can increase maximal payoff
in decision problems carried out via the EWL scheme. Our research
has allowed to formulate Proposition \ref{proposition0} that
points out another peculiarity of quantum games. Unitary
strategies (\ref{operator}) that include classical behavioral ones
(when they are restricted to one-parameter operators) can be
outcome equivalent to unitary operators implementing classical
mixed strategies while behavioral and mixed strategies are not
outcome equivalent in classical decision problems. These
surprising features make quantum games worth further thorough
studies.
\section*{Acknowledgments}
The author is very grateful to his supervisor Prof. J. Pykacz from
the Institute of Mathematics, University of Gda\'nsk, Poland for
his great help in putting this paper into its final form.

\end{document}